\documentclass[12pt]{revtex4}
\usepackage{amssymb}
\usepackage{latexsym}
\usepackage{epsfig}
\begin{document}                             

\title{Does HM mechanism work in string theory?}

\author{Alireza Sepehri$^{1}$  , Somayyeh Shoorvazi$^{2}$\footnote{shoorvazi@gmail.com},Mohammad Ebrahim Zomorrodian $^{3}$ }
\address{$^1$ Faculty of Physics, Shahid Bahonar University, P.O. Box 76175, Kerman, Iran \\$2$ Islamic Azad University,Neyshabur Branch,Neyshabur, Iran\\ $^3$ Department of Physics, Ferdowsi University of Mashhad, 91775-1436, Mashhad, Iran.}

\begin{abstract}
\textbf{Abstract}: The correspondence principle offered a unique opportunity to test the Horowitz and Maldacena mechanism at correspondence point "the centre of mass energies around $(M_{s}/(g_{s})^{2} )$ ".First by using  Horowitz and Maldacena proposal, the black hole final state for closed strings is studied and  the entropy of these states is calculated.Then, to consider the closed string states, a copy of the original Hilbert space is constructed with a set of operators of creation/anihilation that have the same commutation properties as the original ones. The total Hilbert space is the tensor product of the two spaces$ H_{right}\otimes H_{left} $  , where in this case $ H_{left/right}$  denotes the physical quantum states space of the closed string .It is shown that closed string states can be represented by a maximally entangled two-mode squeezed state of the left and right spaces of closed string. Also, the entropy for these  string states is calculated.It is found that black hole entropy matches the  closed string entropy at transition point.This means that our result is consistent with correspondence principle  and thus  HM mechanism in string theory works . Consequently the unitarity of the black hole in string theory can be reconciled.However Gottesman and Preskill point out that, in this scenario, departures from unitarity can arise due to interactions between the collapsing body and the infalling Hawking radiation inside the event horizon and information can be lost. By extending the Gottesman and Preskill method to string theory, the amount of  information transformation from the matter to the state of outgoing Hagedorn radiation for closed strings is obtained.It is observed that information is lost for closed strings.
\end{abstract}

 \maketitle
\section{Introduction}
   Recently Horowitz and Maldacena have suggested a mechanism to reconcile
the unitarity of black hole evaporation
\cite{r1}. In this proposal, there were three
different Hilbert spaces that belong to degrees of freedom of matter, incoming, and
outgoing radiation. The total state of a black hole is a direct product of matter state
and the entangled states of the states inside and outside of the event horizon. When a black hole begins to evaporate, the matter state will be in a maximally entangled
state with incoming Hawking radiation. To describe the unknown effects of
quantum gravity, an additional unitary transformation S is introduced. The outside
state of a black hole is easily obtained by projecting this entangled state on the total
state of the black hole (HM mechanism)\cite{r1}. The question arises is that if there is a
possibility for testing HM mechanism in string theory. Using the correspondence
principle we can examine this idea at transition point.

   The correspondence principle in string theory has provided us with a
convincing picture of the evolution of a black hole at the last stages of its
evaporation \cite{r2}. As the black hole shrinks, it eventually reaches the
"correspondence point" at the centre of mass energies around $(M_{s}/(g_{s})^{2} )$, and makes a
transition to a configuration dominated by a highly excited long string (string
ball). Here $g_{s}$ and $M_{s}$ are the string coupling and string mass respectively. At this
point, the production cross section , entropy and other properties of string balls will
match the production cross section , entropy and other properties of black holes\cite{r2,r3,r4,r5}.We can obtain the final state of black hole by using HM proposal and derive the
entropy for this black hole. Then we will compare this entropy with string ball
entropy at correspondence point. If these entropies are the same ,HM mechanism
works and the unitarity of the black hole can be reconciled. However Gottesman
and Preskill point out that, in this scenario, departures from unitarity can arise due
to interactions between the collapsing body and the infalling Hawking radiation
inside the event horizon. The amount of information loss when a black hole
evaporates depends on the extent to which these interactions are entangling\cite{r5}.

   The important question is that if string model is sufficient to solve the
information-loss problem. Using the Gottesman and Preskill method\cite{r5},we
calculate the information transformation from the matter to the state of outgoing Hawking radiation for closed strings. If information transformation is complete ,we
can solve the information loss paradox in string theory.

   The outline of the paper is as the following. In section \ref{II} we obtain the
entangled states on outside and inside Hilbert spaces of black hole for closed
strings .Then we study the entangled states on inside and matter Hilbert spaces of
the black hole for strings in section \ref{III} . Next we obtain the final state of black
hole for closed strings in section \ref{IV}. We compare the entropy carried out by closed strings with string ball entropy at transition point in section \ref{V}. Finally we
calculate the information transformation from collapsing matter to outside of black
hole in section \ref{VI} .We summarize our results in section
\ref{sum}.

\section{The black hole states for closed string}\label{II}
In this section we discuss the evaporation of Schwarzschild black holes into
closed strings and show that the ground state for these strings is a maximally
entangled state on outside and inside Hilbert spaces of black holes .

First let us consider a bosonic closed string in flat space time and then extend
it to this space time. A closed string moving in a d dimensional flat space-time is
described by d two-dimensional fields $x^{\mu}(\sigma,\tau),\mu=0,1,...,d-1$with $ 0\leq \sigma \leq 2\pi$ and can
be expand in modes as\cite{r6}:
\begin{equation}\label{e1}
X^{\mu}=x_{0}^{\mu}+\sqrt{\acute{2\alpha}}p^{\mu}\tau +i\sqrt{\acute{\alpha}/2}\sum_{n\neq0}\frac{e^{-in\tau}}{n}(\alpha_{n}^{\mu}e^{in\sigma}+
\bar{\alpha}_{n}^{\mu}e^{-in\sigma})
\end{equation}
where $ l_{s}=\sqrt{2\acute{\alpha}}=\frac{1}{\sqrt{\pi T}}$is the fundamental string length and T is string tension. The
center of mass coordinate is $ x_{0}^{\mu}$ and string momentum is $ p^{\mu}$ . The commutation
relations satisfied by the various operators are:\cite{r7}
\begin{equation}\label{e2}
[\bar{\alpha}_{m}^{\mu},\bar{\alpha}_{n}^{\nu}]=m\delta_{m+n,0}g^{\mu\nu},
[\alpha_{m}^{\mu},\alpha_{n}^{\nu}]=m\delta_{m+n,0}g^{\mu\nu} ,[\bar{\alpha}_{m}^{\mu},\alpha_{n}^{\nu}]=0
\end{equation}
where $g^{\mu\nu}$ is the metric in d dimensional space-time.

Now we consider excited closed string fields in black hole space-time .The
stationary four dimensional Schwarzschild black hole with regarding internal space
is represented by the metric \cite{r8}:
\begin{equation}\label{e3}
ds^{2}=-(1-\frac{2M_{BH}}{r})dt^{2}+(1-\frac{2M_{BH}}{r})^{-1}dr^{2}+r^{2}d\Omega^{2}+g_{ij}dx^{i}dx^{j}
\end{equation}
where $M_{BH}$ is the mass of black hole .At $r=2M_{BH}$ ,the Schwarzschild black hole
has an event horizon . Here, we assume a direct product structure with an internal
space parameterized by $x^{i}$, which has the metric $g_{ij}$ which can be taken to be flat.

In the black hole space-time ,the physical degrees of freedom dynamics for
bosonic closed strings is governed by the equation:
\begin{equation}\label{e4}
(-g)^{1/2}\frac{\partial}{\partial x^{\mu}}[g^{\mu\nu}(-g)^{1/2}\frac{\partial}{\partial x^{\nu}}]X^{i}=0,i=1,...,d-1
\end{equation}
for the transversal coordinates of black hole observers, which is also valid for
inertial coordinates. In Kruskal coordinates the metric of black hole becomes\cite{r8}:

\begin{eqnarray}
ds^{2}=-2M_{BH}\frac{e^{-\frac{r}{2M_{BH}}}}{r}d\bar{u}d\bar{v}+r^{2}d\Omega^{2}+g_{ij}dx^{i}dx^{j}\nonumber \\
\bar{u}=-4M_{BH}e^{-u/4M_{BH}},\bar{v}=-4M_{BH}e^{-v/4M_{BH}}\nonumber \\
u=t-r^{\ast},v=t+r^{\ast},r^{\ast}=-r-2M_{BH}\ln{|r-2M_{BH}|}
\label{e5}
\end{eqnarray}
 Since the Killing vector in Kruskal coordinate is given by$ \frac{\partial}{\partial \bar{u}}$
on past horizon "$H^{-}$" , the positive frequency normal mode solution in Kruskal coordinate is
approximated by :
\begin{equation}\label{e6}
\bar{X}^{i}\propto e^{-iE_{s}\bar{u}}
\end{equation}
where $E_{s}$ is the string mass and i=1,..d-1 is the free index that is related to ith component of string. Using this fact that $\bar{v} = 0$ on past horizon "$H^{-}$" \cite{r9} we can estimate the original
positive frequency normal mode on past horizon:
\begin{eqnarray}
X^{i}  & \propto & e^{-iE_{s}u}=(\frac{|\bar{u}|}{4M_{BH}})^{-i4M_{BH}E_{s}}=
\{
\begin{array}{cc}
(-\frac{\bar{u}}{4M_{BH}})^{-i4M_{BH}E_{s}} & \text(region I) \\
(\frac{\bar{u}}{4M_{BH}})^{-i4M_{BH}E_{s}} & \text(region II)
\end{array}\label{e7}
\end{eqnarray}
where $E_{s}$ is the energy of string. In Eq.(\ref{e7}), we can use the fact that
 $(-1)^{-i4M_{BH}E_{s}}=e^{4M_{BH}E_{s}}$. Using equation (\ref{e7}) we observe that the string states in the
horizon satisfy the following condition\cite{r6}:

\begin{eqnarray}
(X^{i}_{out}-\tanh r_{E_{s}}X^{i}_{in})|\text{BH}\rangle^{i}_{in\bigotimes out}=0 \nonumber \\ \tanh r_{E_{s}}=e^{-4M_{BH}E_{s}}\label{e8}
\end{eqnarray}
which actually constitutes a boundary state. In fact, we can view Hawking
radiation as the pair creation of a positive energy string that goes to infinity and a
negative energy string that falls into the black hole. The pair is created in a
particular entangled state. So the Unruh state can be viewed as an entangled
thermal state. Using the expansion in modes for closed strings (equation \ref{e1}) we may
write:
\begin{equation}\label{e9}
 (\alpha^{i}_{n,out}- \tanh r_{E_{s},\lambda}\alpha_{n,in}^{i\dag})|\text{BH,n}\rangle^{i}_{in\bigotimes out}=0 \text{for n}\neq0
\end{equation}
\begin{equation}\label{e10}
 (\bar{\alpha}^{i}_{n,out}- \tanh r_{E_{s},-\lambda}\bar{\alpha}_{n,in}^{i\dag})|\text{BH,n}\rangle^{i}_{in\bigotimes out}=0 \text{for n}\neq0
\end{equation}
\begin{equation}\label{e11}
\tanh r_{E_{s},\lambda}=e^{-4M_{BH}E_{s}-2in\lambda}
\end{equation}
\begin{equation}\label{e12}
(x^{i}_{0,out}-x^{i}_{0,in})|\text{BH,0}\rangle^{i}_{in\bigotimes out}=0\text{for n}=0
\end{equation}
where we have defined the respective string coordinates up to a shift:$\sigma_{out}\equiv \sigma_{in}+\lambda $\cite{r6,r10}.
Now, we assume that the Kruskal vacuum $|\text{BH}\rangle^{i}_{in\bigotimes out}$ is related to the flat  vacuum $|0\rangle_{s}$ by
\begin{equation}\label{e13}
|\text{BH}\rangle^{i}_{in\bigotimes out}=F_{1}(\alpha^{i}_{n,out},\alpha^{\dag i}_{n,in})F_{2}({\bar{\alpha}^{i}_{n,out},\bar{\alpha}^{\dag i}_{n,in}})|0\rangle_{s}
\end{equation}
where $F_{1}$ and $F_{2}$ are some functions of closed string creation-annihilation
operators to be determined later. These operators act on black hole vacuum and
form string modes inside and outside of event horizon. It is clear that when the
string does not intersect the horizon, the F transformations are trivial.

From $[\alpha^{i}_{n,out},\alpha^{\dag i}_{n,out}]=1$,we obtain $[\alpha^{i}_{n,out},(\alpha^{\dag i}_{n,out})^{m}]=\frac{\partial}{\partial \alpha^{\dag i}_{n,out} }(\alpha^{\dag i}_{n,out})^{m}$ and$[\alpha^{i}_{n,out},F]=\frac{\partial}{\partial \alpha^{\dag i}_{n,out} }F$
\cite{r11}.Then using Eqs. (\ref{e9})and (\ref{e10}), we get the following differential equations for $F_{1}$  and $F_{2}$ .
\begin{equation}\label{e14}
 (\frac{\partial F_{1}}{\partial \alpha^{\dag i}_{n,out}}-tanh r_{E_{s},\lambda}\alpha^{\dag i}_{n,in}F)=0
\end{equation}
\begin{equation}\label{e15}
 (\frac{\partial F_{2}}{\partial \bar{\alpha}^{\dag i}_{n,out}}-tanh r_{E_{s},-\lambda}\bar{\alpha}^{\dag i}_{n,in}F)=0
\end{equation}
The solution is given by
\begin{equation}\label{e16}
 F_{1}=e^{tanh r_{E_{s},\lambda}\alpha^{\dag i}_{n,out}\alpha^{\dag i}_{n,in}}, F_{2}=e^{tanh r_{E_{s},-\lambda}\bar{\alpha}^{\dag i}_{n,out}\bar{\alpha}^{\dag i}_{n,in}}
\end{equation}
By substituting Eq.(\ref{e16}) into Eq. (\ref{e13}) and by properly normalizing the state vector, we get
\begin{eqnarray}|\text{BH,n}\rangle^{i}_{in\bigotimes out}
=  N_{BH,n} e^{tanh r_{E_{s},\lambda}\alpha^{\dag i}_{n,out}\alpha^{\dag i}_{n,in}}e^{tanh r_{E_{s},-\lambda}\bar{\alpha}^{\dag i}_{n,out}\bar{\alpha}^{\dag i}_{n,in}} |0\rangle_{s} \nonumber \\
= \frac{1}{coshr_{E_{s},\lambda}}\frac{1}{coshr_{E_{s},-\lambda}}  \sum_{m,\bar{m}}tanh^{m}r_{E_{s},\lambda}tanh^{\bar{m}}r_{E_{s},-\lambda}|n,m,\bar{m}\rangle_{in}\otimes|n,m,\bar{m}\rangle_{out}\label{e17}
\end{eqnarray}
where $|n,m,\bar{m}\rangle_{in}$  and $|n,m,\bar{m}\rangle_{out}$  are orthonormal bases (normal mode solutions) for mode n that act on $H_{in}$ and $H_{out}$ respectively . $N_{BH,n}$  is the normalization constant.The total ground state for black hole can be derived as:
\begin{eqnarray}|\text{BH}\rangle^{i}_{in\bigotimes out}
=  \prod_{n} \frac{1}{cosh^{n}r_{E_{s},\lambda}}\frac{1}{cosh^{n}r_{E_{s},-\lambda}}  \sum_{m,\bar{m}}tanh^{nm}r_{E_{s},\lambda}tanh^{n\bar{m}}r_{E_{s},-\lambda}\otimes\nonumber \\|n,m,\bar{m}\rangle_{in}\otimes|n,m,\bar{m}\rangle_{out}\label{e18}
\end{eqnarray}
     Eq. (\ref{e18}) states that the black hole state decomposes to many entangled
states with different energies for closed string . On the other hand, this equation
indicates that non-local physics would be required to transmit the string information
outside the black hole, and that inside and outside the Hilbert spaces are
dependent.
\section{The evaporation of black hole to closed strings}\label{III}
Previously it has been investigated that the field inside the event horizon can be
decomposed into the collapsing matter field and that the advanced wave incoming
from infinity has the same form as the Hawking radiation\cite{r9,r12}.We extend these
calculations to string theory for closed strings .We introduce the advanced
collapsing shell metric\cite{r9} in d dimensions at correspondence point as:
\begin{eqnarray}
ds^{2}  & = &
\{
\begin{array}{cc}
-d\acute{\tau}^{2}+dr^{2}+r^{2}d\Omega^{2}+g_{ij}dx^{i}dx^{j} & r<R(\acute{\tau})\\
-(1-\frac{2M_{BH}}{r})dt^{2}+(1-\frac{2M_{BH}}{r})^{-1}dr^{2}+r^{2}d\Omega^{2}+g_{ij}dx^{i}dx^{j}& r>R(\acute{\tau})
\end{array}\label{e19}
\end{eqnarray}

The shell radius $R(\acute{\tau})$ is defined by\cite{r9}:
\begin{eqnarray}
R(\acute{\tau})  & = &
\{
\begin{array}{cc}
R_{0} & \acute{\tau}<0\\
R_{0}-v\acute{\tau}& \acute{\tau}>0
\end{array}\label{e20}
\end{eqnarray}
where $ \acute{\tau}$ is the proper time. The advanced coordinates has been defined as\cite{r9,r12,r13}:
\begin{equation}\label{e21}
V=\acute{\tau}+r-R_{0},U=\acute{\tau}-r+R_{0}
\end{equation}
Now we consider the closed string modes inside the black hole (on future horizon" $H^{+}$" ), which is
given by\cite{r9}:
\begin{eqnarray}
X^{+i}  & = &
\{
\begin{array}{cc}
(-1)^{-i4M_{BH}M_{s}}(1-\frac{\nu V}{(1-\nu)(R_{0}-2M_{BH})})^{-i4M_{BH}E_{s}} & V>\frac{(1-\nu)(R_{0}-2M_{BH})}{\nu}\text{matter wave function} \\
(1-\frac{\nu V}{(1-\nu)(R_{0}-2M_{BH})})^{-i4M_{BH}E_{s}} & V<\frac{(1-\nu)(R_{0}-2M_{BH})}{\nu}\text{inside wave function}
\end{array}\label{e22}
\end{eqnarray}
Using Eq. (\ref{e22}) we observe that the string states in the horizon satisfy the
following condition:
\begin{eqnarray}
(X^{i}_{matter}-\tanh r_{E_{s}}X^{i}_{in})|\text{BH}\rangle^{i}_{in\bigotimes out}=0 \nonumber \\ \tanh r_{E_{s}}=e^{-4M_{BH}E_{s}}\label{e23}
\end{eqnarray}
which actually constitutes a boundary state. Using the expansion in modes for
closed strings (equation (\ref{e23})) we write:
\begin{equation}\label{e24}
 (\alpha^{i}_{n,matter}- \tanh r_{E_{s},\lambda}\alpha_{n,in}^{i\dag})|\text{BH,n}\rangle^{i}_{in\bigotimes out}=0 \text{for n}\neq0
\end{equation}
\begin{equation}\label{e25}
 (\bar{\alpha}^{i}_{n,matter}- \tanh r_{E_{s},-\lambda}\bar{\alpha}_{n,in}^{i\dag})|\text{BH,n}\rangle^{i}_{in\bigotimes out}=0 \text{for n}\neq0
\end{equation}
\begin{equation}\label{e26}
\tanh r_{E_{s},\lambda}=e^{-4M_{BH}E_{s}-2in\lambda}
\end{equation}
\begin{equation}\label{e27}
(x^{i}_{0,matter}-x^{i}_{0,in})|\text{BH,0}\rangle^{i}_{in\bigotimes out}=0 \text{for n}=0
\end{equation}
The solution of equations (\ref{e24}) and (\ref{e25} ) may be expressed as:
\begin{eqnarray}|\text{BH,n}\rangle^{i}_{matter\bigotimes in}
=  N_{BH,n} e^{tanh r_{E_{s},\lambda}\alpha^{\dag i}_{n,matter}\alpha^{\dag i}_{n,in}}e^{tanh r_{E_{s},-\lambda}\bar{\alpha}^{\dag i}_{n,matter}\bar{\alpha}^{\dag i}_{n,in}} |0\rangle_{s} \nonumber \\
= \frac{1}{coshr_{E_{s},\lambda}}\frac{1}{coshr_{E_{s},-\lambda}}  \sum_{m,\bar{m}}tanh^{m}r_{E_{s},\lambda}tanh^{\bar{m}}r_{E_{s},-\lambda}|n,m,\bar{m}\rangle_{matter}\otimes|n,m,\bar{m}\rangle_{in}\label{e28}
\end{eqnarray}
To obtain the total ground state inside the black hole ,we multiply the entangled
states for different modes of closed string:
\begin{eqnarray}|\text{BH}\rangle^{i}_{matter\bigotimes in}
=  \prod_{n} \frac{1}{cosh^{n}r_{E_{s},\lambda}}\frac{1}{cosh^{n}r_{E_{s},-\lambda}}  \sum_{m,\bar{m}}tanh^{nm}r_{E_{s},\lambda}tanh^{n\bar{m}}r_{E_{s},-\lambda}\otimes\nonumber \\|n,m,\bar{m}\rangle_{matter}\otimes|n,m,\bar{m}\rangle_{in}\label{e29}
\end{eqnarray}
This equation shows that matter and inside states of closed strings in evaporating
black holes are entangled .This entanglement decreases for massive black holes
and high energy strings.

\section{The black hole final state for closed strings}\label{IV}
Now we extend the Horowitz and Maldacena mechanism to closed strings.
As mentioned above ,before black hole evaporation, there is an entanglement
between the inside and outside of the horizon, but matter itself is not entangled.
\begin{equation}\label{e30}
|\text{BH}\rangle^{i}_{matter\bigotimes in \bigotimes out}=|\text{BH}\rangle^{i}_{matter}\otimes|\text{BH}\rangle^{i}_{ in \bigotimes out}
\end{equation}
After evaporation, we can describe the state of the black hole as an entangled state
of both matter and the inside of the black hole:
\begin{equation}\label{e31}
^{i}_{matter\otimes in}\langle B\acute{H}|=^{i}_{matter\otimes in}\langle BH|(S\otimes I)
\end{equation}
To describe the unknown effects of quantum gravity, an additional unitary
transformation S is introduced\cite{r1,r9,r12}.The outside state of black hole can be
described as:
\begin{eqnarray}
|\text{BH}\rangle^{i}_{out}=^{i}_{matter\otimes in}\langle BH|(S\otimes I)|\text{BH}\rangle^{i}_{matter\bigotimes in \bigotimes out}=\nonumber \\
\prod_{n} \frac{1}{cosh^{2n}r_{E_{s},\lambda}}\frac{1}{cosh^{2n}r_{E_{s},-\lambda}}  \sum_{m,\bar{m}}tanh^{2nm}r_{E_{s},\lambda}tanh^{2n\bar{m}}r_{E_{s},-\lambda}\otimes\nonumber \\
 _{matter}\langle n,m,\bar{m}|S|BH\rangle_{matter}\otimes|n,m,\bar{m}\rangle_{out}
\label{e32}
\end{eqnarray}
Until now, we have shown that the internal stationary state of the black hole
for closed strings can be represented by a maximally entangled two-mode squeezed state of collapsing matter and infalling string radiation. The outgoing string
radiation is obtained by the final state projection on the total wave function, which
looks like a quantum teleportation process without the classical information
transmitted. The black hole evaporation process as observed by the observer
outside the event horizon is a unitary process but non-locality is required to
transmit the string outside the event horizon. The uniqueness of the final state for
closed strings seems to prevent any information from getting "stuck" there, so that
all of the string information encoded in the initial collapsing black hole can appear
in the outgoing string radiation. However Gottesman and PresKill believe that
interactions between the incoming Hawking radiation and the collapsing matter
can spoil the unitary nature of the black-hole evaporation \cite{r5}, destroying some or
all of the quantum information inside the hole \cite{r14}.

\section{HM mechanism at correspondence point}\label{V}
After the first emission, the mass of the black hole is reduced to $M_{BH}-E_{s}$ due to
energy conservation .Consequently after n emissions the mass of black hole is
reduced to $M_{BH}-nE_{s}$ .According to the string theory of quantum gravity, the
minimum mass above which a black hole can be treated general relativistically is
\cite{r2,r15}:
\begin{equation}\label{e33}
M_{min}\sim\frac{M_{s}}{g_{s}^{2}}
\end{equation}
The properties of a black hole with mass $M_{min}$ matches those of a string ball with
the same mass. This is called the correspondence principle and the mass at which
this happens is the correspondence point \cite{r2,r3,r4}. At this point that black hole makes
a transition to a string, it can become a single string, multiple string, or
radiation\cite{r2}. At this point , the entropy of black hole and excited string should
be the same\cite{r16}.
At this stage we intend to work with entropy, to compare final states in black
holes as well as in string balls at corresponding point. First we write down the
density matrices $\rho $ for the closed strings, using the direct product of
$ |BH\rangle_{out}$and $ _{out}\langle BH|$ . This gives the density matrices for closed strings :
\begin{eqnarray}
\rho=|BH\rangle_{out}\langle BH |= \nonumber \\
\prod_{n}\frac{1}{cosh^{4n}r_{E_{s},\lambda}}\frac{1}{cosh^{4n}r_{E_{s},-\lambda}}\otimes \nonumber \\ \sum_{m,\bar{m}}e^{-16\pi M_{BH}nmE_{s}}e^{-16\pi M_{BH}n\bar{m}E_{s}}|n,m,\bar{m}\rangle_{out}\langle n,m,\bar{m} |
\label{e34}
\end{eqnarray}
We calculate the entropy S of the quantum states using the prescription of von
Neumann \cite{r17}, which makes direct use of the density matrix $\rho $ . It is simply written
as:
\begin{equation}\label{e35}
S=-Tr(\rho\ln{\rho})
\end{equation}
where the trace "Tr" is taken over the quantum states. We now apply these
definitions to the entropy for the closed strings. After some mathematical
manipulations for high energy strings,we obtain:
\begin{equation}\label{e36}
S_{BH}=-\sum_{n=1}sinh^{2}r_{E_{s},\lambda}\ln{sinh^{2}r_{E_{s},\lambda}}-\sum_{n=1}sinh^{2}r_{E_{s},-\lambda}\ln{sinh^{2}r_{E_{s},-\lambda}}
\end{equation}
\begin{equation}\label{e37}
sinh^{2}r_{E_{s},\lambda}=\frac{e^{-4\pi M_{BH}E_{s}-2n\lambda}}{1-e^{-4\pi M_{BH}E_{s}-2in\lambda}}=\frac{e^{-\frac{E_{s}}{T_{BH}}-2in\lambda}}{1-e^{-\frac{E_{s}}{T_{BH}}-2in\lambda}}
\end{equation}
For$ \lambda=0$ the black hole entropy can be written as:

\begin{equation}\label{e38}
S_{BH}=-2\sum_{n=1}sinh^{2}r_{E_{s}}\ln{sinh^{2}r_{E_{s}}}
\end{equation}
where the Hawking temperature is given by\cite{r18}:
\begin{equation}\label{e39}
T_{BH}=\frac{1}{4\pi M_{BH}}
\end{equation}
To consider the closed string states, a copy of the original Hilbert space may be
constructed with a set of operators of creation/anihilation that have the same
commutation properties as the original ones. The total Hilbert space is the tensor product of the two spaces $H_{right}$ ,$H_{left}$ , where in this case $H_{left / right}$  denotes the
physical quantum states space of the closed string .The left hand of string affects
the right hand of string by entanglement\cite{r6,r10}.Thus, in agreement with Refs.
\cite{r6,r10} we may write the Bogoliubov transformation between the Minkowski and
string creation and annihilation operators :
\begin{equation}\label{e40}
(coshr_{T_{sb}}\alpha_{n}^{i}-sinhr_{T_{sb}}\bar{\alpha}_{n}^{\dag i})|\text{closed string,n}\rangle=0
\end{equation}
\begin{equation}\label{e41}
(coshr_{T_{sb}}\bar{\alpha}_{n}^{i}-sinhr_{T_{sb}}\alpha_{n}^{\dag i})|\text{closed string,n}\rangle=0
\end{equation}
\begin{equation}\label{e42}
tanhr_{T_{sb}}=e^{\frac{-E_{s}}{T_{sb}}}
\end{equation}
The Hagedorn temperature of an excited string matches the Hawking temperature
of a black hole at the correspondence point. The Hagedorn temperature is given
by\cite{r2,r3,r4}:
\begin{equation}\label{e43}
T_{sb}=\frac{M_{s}}{\sqrt{8\pi}}
\end{equation}
We can rewrite equations (\ref{e40})and(\ref{e41} )as the following:
\begin{equation}\label{e44}
(coshr_{T_{sb}}\frac{\partial}{\partial\alpha_{n}^{\dag i}}-sinhr_{T_{sb}}\bar{\alpha}_{n}^{\dag i})|\text{closed string,n}\rangle=0
\end{equation}
\begin{equation}\label{e45}
(coshr_{T_{sb}}\frac{\partial}{\partial\bar{\alpha}_{n}^{\dag i}}-sinhr_{T_{sb}}\alpha_{n}^{\dag i})|\text{closed string,n}\rangle=0
\end{equation}
The solution of these equations is given by:
\begin{equation}\label{e46}
|\text{closed string,n}\rangle=\frac{1}{cosh^{2}r_{T_{sb}}}\sum_{m,\bar{m}=0}^{\infty}tanh^{m+\bar{m}}r_{T_{sb}}|m\rangle\otimes|\bar{m}\rangle
\end{equation}
Thus we can calculate the total ground state for closed strings as :
\begin{equation}\label{e47}
|\text{closed string}\rangle=\prod_{n}\otimes|\text{closed string,n}\rangle
\end{equation}
The closed string entropy can be calculated as :
\begin{eqnarray}
S_{sb}=-\sum_{n=1}\alpha_{n}^{i\dag}\alpha_{n}^{i}\ln{sinh^{2}r_{T_{sb}}}-\sum_{n=1}\alpha_{n}^{i}\alpha_{n}^{i\dag}\ln{cosh^{2}r_{T_{sb}}}\nonumber \\-\sum_{n=1}\bar{\alpha}_{n}^{i\dag}\bar{\alpha}_{n}^{i}\ln{sinh^{2}r_{T_{sb}}}-\sum_{n=1}\bar{\alpha}_{n}^{i}\bar{\alpha}_{n}^{i\dag}\ln{cosh^{2}r_{T_{sb}}}
\label{e48}
\end{eqnarray}
where
\begin{equation}\label{e49}
sinh^{2}r_{T_{sb}}=\frac{e^{\frac{E_{s}}{T_{sb}}}}{1-e^{\frac{E_{s}}{T_{sb}}}}
\end{equation}
By setting $T_{SB}=T_{H}$ ,for high energy strings, the entropy operator is obtained as :
\begin{eqnarray}
S_{sb}=-\sum_{n=1}\alpha_{n}^{i\dag}\alpha_{n}^{i}\ln{sinh^{2}r_{E_{s},\lambda=0}}-\sum_{n=1}\alpha_{n}^{i}\alpha_{n}^{i\dag}\ln{cosh^{2}r_{E_{s},\lambda=0}}\nonumber \\-\sum_{n=1}\bar{\alpha}_{n}^{i\dag}\bar{\alpha}_{n}^{i}\ln{sinh^{2}r_{E_{s},\lambda=0}}-\sum_{n=1}\bar{\alpha}_{n}^{i}\bar{\alpha}_{n}^{i\dag}\ln{cosh^{2}r_{E_{s},\lambda=0}}
\label{e50}
\end{eqnarray}
To obtain the expectation value of closed excited string entropy, we calculate the
expectation value of annihilation/creation operators of closed strings:
\begin{eqnarray}
\langle closed string,n |\alpha_{n}^{ i\dag}\alpha_{n}^{ i}|closed string,n\rangle=\nonumber \\\langle closed string,n |\bar{\alpha}_{n}^{i\dag }\bar{\alpha}_{n}^{ i}|closed string,n\rangle=sinh^{2}r_{E_{s}}
\label{e51}
\end{eqnarray}
Now we can obtain the expectation value of string entropy as following:
\begin{equation}\label{e52}
\langle S_{sb}\rangle=-2\sum_{n=1}sinh^{2}r_{E_{s}}\ln({sinh^{2}r_{E_{s}}})=S_{BH}
\end{equation}
This entropy is approximately equal to the black hole entropy with the same mass.

The correspondence principle offered a unique opportunity to test the Horowitz
and Maldacena mechanism by calculating the black hole entropy at corresponding
point and comparing it with the string ball entropy. We calculated the final outside
state of black hole by applying Horowitz and Maldacena idea and obtained the
entropy that carried out by these states. We found that this entropy matches the
string ball entropy at corresponding point. This means that our result is consistent
with correspondence principle and thus Horowitz and Maldacena mechanism
works.
\section{Information loss for closed string in black holes} \label{VI}
The information loss in a black hole has been a serious challenge to modern
physics, because it contradicts the basic principles of quantum mechanics. This
argument predicts that a process of black hole formation and evaporation is not
unitary. Gottesman and Preskill explained how interactions between the collapsing
body and infalling Hawking radiation inside the event horizon could cause a loss of
information. Gottesman and Preskill introduce a unitary transformation U to
describe interactions between matter and the inside of the horizon \cite{r5}. In their
fanciful language, the transformation U can be interpreted as an interaction
between the information's past and future. They obtained the amount of this
information loss for particles \cite{r5}.Using their method we calculate the information
transformation from collapsing matter to outside of black hole as:
\begin{eqnarray}
T_{\text{closed string}}=_{matter\otimes in}\langle B\acute{H}|BH\rangle_{in\otimes out}= \nonumber \\\prod_{n} \frac{1}{cosh^{2n}r_{E_{s},\lambda}}\frac{1}{cosh^{2n}r_{E_{s},-\lambda}}  \sum_{m,\bar{m}}tanh^{2nm}r_{E_{s},\lambda}tanh^{2nm}r_{E_{s},-\lambda}\otimes\nonumber \\
_{matter}\langle m,\bar{m},n|_{in}\langle m,\bar{m},n|S\otimes U|m,\bar{m},n\rangle_{in}|m,\bar{m},n\rangle_{out}
\label{e53}
\end{eqnarray}

\begin{eqnarray}
f=|T_{\text{closed string}}|^{2}=\prod_{n} |\frac{1}{cosh^{4n}r_{E_{s},\lambda}}\frac{1}{cosh^{4n}r_{E_{s},-\lambda}}|\otimes\nonumber \\  \sum_{m,\bar{m},\acute{m},\acute{\bar{m}}}e^{-8\pi n M_{BH}\omega_{n}(m+\acute{m})}e^{-8\pi n M_{BH}\omega_{n}(\bar{m}+\acute{\bar{m}})}\otimes\nonumber \\_{out}\langle \acute{m},\acute{\bar{m}},n|_{in}\langle \acute{m},\acute{\bar{m}},n|_{matter}\langle m,\bar{m},n|_{in}\langle m,\bar{m},n|SS^{\dag}\otimes\nonumber \\  UU^{\dag}|m,\bar{m},n\rangle_{in}|m,\bar{m},n\rangle_{out}
|\acute{m},\acute{\bar{m}},n\rangle_{in}|\acute{m},\acute{\bar{m}},n\rangle_{matter}
=\nonumber \\\prod_{n} |\frac{1}{cosh^{4n}r_{E_{s},\lambda}}\frac{1}{cosh^{4n}r_{E_{s},-\lambda}}|  \sum_{m,\bar{m}}e^{-16\pi n M_{BH}E_{s}m}e^{-16\pi n M_{BH}E_{s}\bar{m}}=\nonumber \\ \prod_{n}(\frac{1+e^{-16\pi M_{BH}E_{s}}-2e^{-8\pi M_{BH}E_{s}}\cos{2n\lambda}}{1-e^{-16\pi M_{BH}E_{s}}})^{2n}
\label{e53}
\end{eqnarray}
\begin{equation}\label{e54}
SS^{\dag}=I,UU^{\dag}=I
\end{equation}

If information transformation from the collapsing matter to the state of
outgoing Hawking radiation is complete, then the value of f should be one \cite{r5,r14}.
Thus, we observed that due to the deviation of the f from unity, some of
information is lost. It seems that for all finite values of $E_{s}$ , the value of f is not unity
and so all information from all string emission processes experiences some degree
of loss.

The purpose of these calculations is to examine the robustness of the escape of
closed strings during black hole evaporation in final state projection models. In
particular, we show that due to interactions between the matter and incoming
Hawking radiation, all closed strings couldn't escape from the hole. Of strings that
escape from the hole, low energy strings are lost, regardless of the number of
strings in the hole to begin with. More precisely, the states of the strings that formed the hole are preserved under black hole evaporation with $ f\simeq1$for high
energy strings and with $ f\simeq0$ for low energy strings. These are the fidelity of
escape of the string states of the collapsing matter: individual strings escape with a
fidelity that approaches one as the energy of string in the hole becomes large.
Since the fidelity is above the threshold for the use of quantum-error correcting
codes, properly high energy strings can escape from the hole with fidelity
arbitrarily close to 1.
\section{Summary and Discussion} \label{sum}
In this manuscript, we discuss the information loss in Schwarzschild black
holes for closed strings. We calculate the amount of information transformation
from collapsing matter to outside of black hole by extending the Gottesman and
Preskil method to string theory .We also test the HM mechanism at correspondence
point. Using this idea we obtain the final state of black hole and calculate the
entropy of these states. Then we compare this entropy with string ball entropy and
observe that these entropies are the same at transition point and thus HM proposal
works.
\acknowledgments{ This work has funded by the vice president for research  Technology of  ferdowsi  University of  Mashhad , Code:2/18709.This work has also  been supported  by the faculty of physics, Shahid  Bahonar University.}

\end{document}